\documentclass[aps,prb,preprint,superscriptaddress,showpacs]{revtex4}

\usepackage{graphicx}
\usepackage{bm}
\usepackage{amsmath}
\usepackage{amssymb}

\begin{document}
\title{Electric charge enhancements in carbon nanotubes :\\
Theory and experiments}

\author{Zhao Wang}
\email{wzzhao@yahoo.fr}
\affiliation{Institute UTINAM, UMR 6213, University of Franche-Comt\'{e}, 25030 Besan\c{c}on Cedex, France.}

\author{Mariusz Zdrojek}
\affiliation{Faculty of Physics, Warsaw University of Technology, Koszykowa 75, 00-662 Warsaw, Poland}

\author{Thierry M\'{e}lin}
\affiliation{D\'epartement ISEN, Institut d'Electronique de Micro\'{e}lectronique et de Nanotechnologie, CNRS-UMR 8520, Avenue Poincar\'{e},BP 60069, 59652 Villeneuve d'Ascq Cedex, France}

\author{Michel Devel}
\affiliation{Institute UTINAM, UMR 6213, University of Franche-Comt\'{e}, 25030 Besan\c {c}on Cedex, France.}

\begin{abstract}
We present a detailed study of the static enhancement effects of electric charges in $\mu$m-long single-walled carbon nanotubes,
using theoretically an atomic charge-dipole model and experimentally electrostatic force microscopy.
We demonstrate that nanotubes exhibit at their ends surprisingly weak charge enhancements which decrease with the nanotube length and increase
with the nanotube radius. A quantitative agreement is obtained between theory and experiments.
\end{abstract}


\maketitle
Understanding of the properties of electric charges in carbon nanotubes (CNTs) is one of the important issues for their promising applications
in nanoelectromechanical systems,\cite{Anantram2006} field emission,\cite{De1995} chemical sensors\cite{Snow2005} and charge storage.\cite{An2001, Cui2002, Ryu2007}
A key-aspect of the electrostatics of these one-dimensional systems is the knowledge of the distribution of electric charges along the nanotubes, because charges
are likely to accumulate at the nanotube ends due to Coulomb repulsion. Theoretical predictions have been established for this effect, but not in the range
of lengths accessible from experiments, so that no comparison has been established between theory and experimental observations so far. More precisely,
the electrostatic properties of single-walled nanotubes (SWCNTs) have been addressed on the one hand using electric force microscopy (EFM)
experiments\cite{Bockrath2002,Jespersen2007} coupled to charge injection techniques\cite{Paillet2005, Zdrojek2006}.
Results obtained for $\mu$m-long nanotubes indicated that electric charges are distributed rather uniformly along the tube length, with however
no theoretical support in this range of nanotube length. On the other hand, in theoretical studies,
density functional theory\cite{Keblinski2002} and classical electrostatics\cite{Li2006a}
calculations have been performed to compute the charge distribution in SWCNTs, and have predicted $U$-like shapes due to a charge accumulation at the nanotube ends. These calculations however only hold for short ($<100$ nm) nanotube lengths, which are not easily accessible from experiments.

It is the scope of this paper to provide a combined experimental and theory work on this issue. We present a detailed study of the static enhancement
effects of electric charges in SWCNTs, using theoretically an atomic charge-dipole model and experimentally electrostatic force microscopy.
It is demonstrated that the U-like shape of the charge distribution expected for short nanotubes is replaced in the case of $\mu$m-long tubes by
weak charge enhancements localized at the nanotube ends, in agreement with the experimental values for the enhancement factors (up to few tens of \%) observed from
EFM and charge injection experiments. The dependence of the charge enhancement factors on the nanotube radius has also been measured from EFM experiments, and falls
in quantitative agreement with theoretical preditions for $\mu$m-long tubes.

The paper is organized as follows : we first describe the numerical calculations of the charge distribution along nanotubes using the atomic charge-dipole model and
the results obtained for short nanotubes with open or closed caps, either considered in vacuum or on a SiO$_2$ substrate. The extrapolation procedure to the case of
$\mu$m-long nanotubes is then presented, and compared with experimental EFM measurements of charge enhancement factors on SWCNTs. We finally discuss the dependence
of the enhancement factors as a function of the nanotube radius. 

In the theoretical calculations presented throughout this work, the interactions between the electric charges and the induced dipoles are described
using the Gaussian-regularized atomic
charge-dipole interaction model,\cite{mayer-07-01, Mayer2008} in which the atoms are treated as interacting polarizable points with free charges,
and the distribution of charges and dipoles are determined by the fact that their static equilibrium state should correspond to the minimum value of
the total molecular electrostatic energy. Compared with classical Coulomb-law-based models in which only charges are considered, this model provides a more
accurate description of electrostatic properties of CNTs, since the charges, the induced dipoles and the atomic polarizabilities are taken into account. 

In order to achieve a valid comparison between experimental data and calculation results, the effect of a SiO$_2$ substrate (nanotubes are usually
deposited on a SiO$_2$ thin film in experiments) is also taken into account in our calculations,
by adding surface-induced terms to the vacuum electrostatic interaction tensors using the method of mirror images.\cite{Jacksonbook1975}
The dielectric constant of SiO$_{2}$ is taken as $4.0$.
The average distance between the bottom of the tubes and the SiO$_{2}$ surface is set to $d = 0.34$ nm after the computed CNT-SiO$_{2}$
long-range interacting configurations from Refs.\cite{Tsetseris2006, Wojdel2005}. Furthermore, we note that $d$ can slightly vary with the tube radius $R$. It will
however be fixed to $0.34$ nm in this work as an average value. The atomic structure of CNTs is then optimized by energy minimization using the method
of conjugated gradient based on a many-body chemical potential model AIREBO (adaptive interatomic reactive empirical bond order).\cite{stuart-00}
Linear charge densities of 0.055$e$/nm have been used in calculations, so as to match linear charge densities observed experimentally.\cite{Zdrojek2008}

\begin{figure}[ht]
\centerline{\includegraphics[width=12cm]{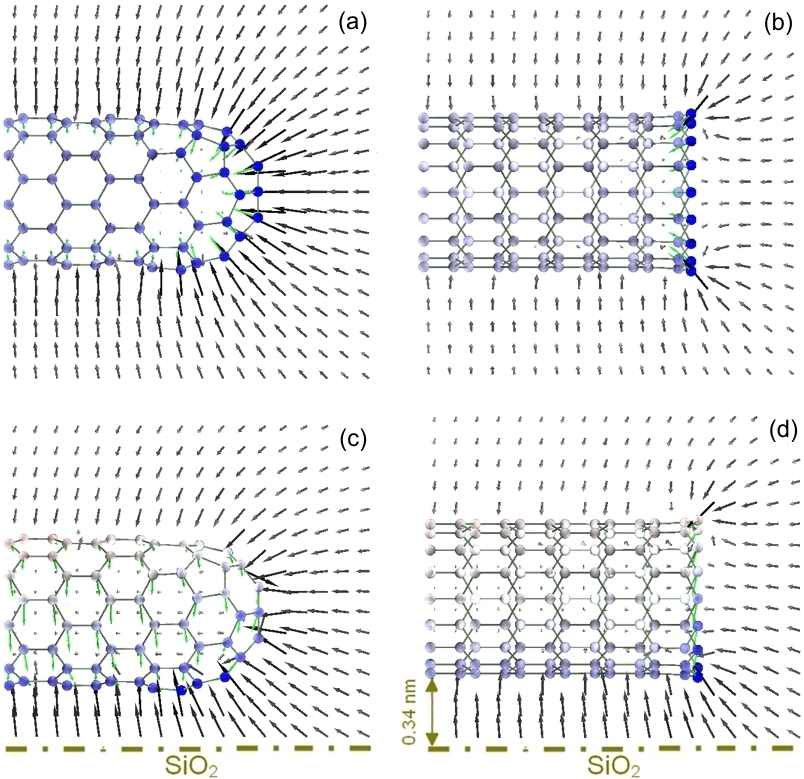}}
\caption{\label{fig:map90tip}
(Color online) Charge distribution at the ends of an open-ended and a closed-ended (9, 0) SWCNTs ($L\approx11.5$ nm, $\sigma^{ave}=-6.6 \times 10^{-4} e/$atom)
in free space ((a) and (b)) and on the SiO$_2$ substrate((c) and (d)). The color of the atoms is proportional to the local charge density. The green vectors stand for the induced atomic dipoles. The dark arrows stand for the local electric fields induced by the net charge, their length and color are proportional to the field intensity. (a) The minimum and maximum atomic charge densities in this tube are: $\sigma^{min}=-5.3 \times 10^{-4}$ $e/$atom and $\sigma^{max}=-16 \times 10^{-4}$ $e/$atom, respectively. (b) $\sigma^{min}=-5.2 \times 10^{-4}$ $e/$atom and $\sigma^{max}=-34 \times 10^{-4}$ $e/$atom. (c) $\sigma^{min}=+5.6 \times 10^{-4}$ $e$/atom and $\sigma^{max}=-34 \times 10^{-4}$ $e$/atom. (d) $\sigma^{min}=+5.8 \times 10^{-4}$ $e$/atom and $\sigma^{max}=-74 \times 10^{-4}$ $e$/atom.}
\end{figure}

To illustrate the typical outputs of the atomic-scale calculations, we show in Fig.\,\ref{fig:map90tip} the charge distribution at the end of a (9, 0) CNT of length $L\approx11.5$ nm and average charge density $\sigma^{ave}=-6.6 \times 10^{-4} e/$atom. The color of the atoms is proportional to their charge in the figure. We represented here for sake of clarity the four distinct situations in which the nanotube exhibits either a closed (Fig.\,\ref{fig:map90tip}a and c) or an open (Fig.\,\ref{fig:map90tip}b and d) cap structure, and the tube is either considered in vacuum (Fig.\,\ref{fig:map90tip}a and b) or deposited on a SiO$_2$ substrate (Fig.\,\ref{fig:map90tip}c and d). As seen from Fig.\,\ref{fig:map90tip}, the charge density at the tube ends is higher than that at other parts of the tubes in all situations. The maximum charge density on the opened cap is here about twice that on the closed one for this small-radius tube. Finally, when the tube is deposited on the SiO$_2$ surface, electrons are attracted by their image charge towards the SiO$_2$ surface,
as a typical semi-space effect.

\begin{figure}[ht]
\centerline{\includegraphics[width=10cm]{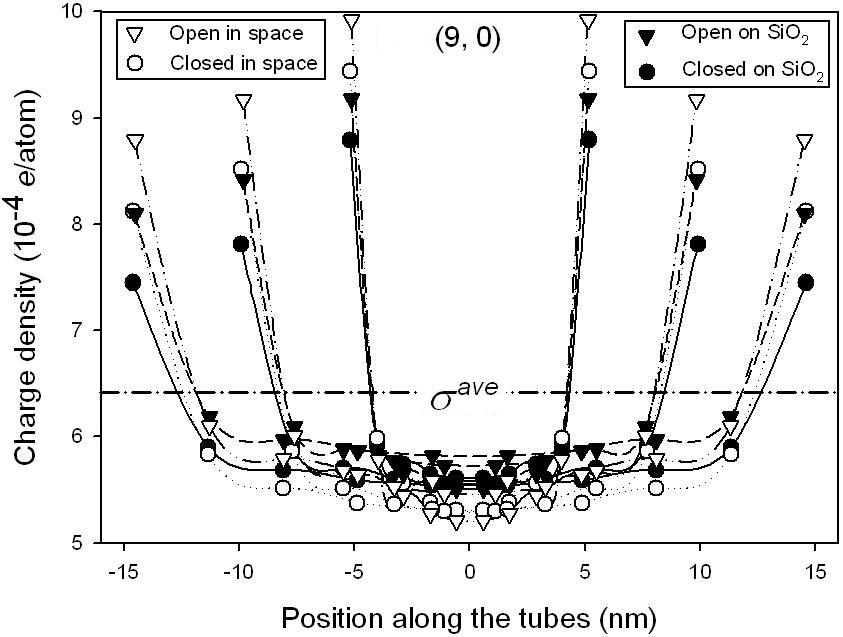}}
\caption{\label{fig:sigma6length}
Charge profile along three (9, 0) SWCNTs with different tube lengths $L$, in space (hollow symbols) and upon a SiO$_2$ surface (solid symbols), using a separation distance $d = 0.34$ nm (see text). The total net charge density on each tube $\sigma^{ave}$ is fixed to $6.4 \times 10^{-4} e/$atom (equivalently, $0.055 e/$ nm). Each point is calculated as the average value of the charge carried by the nanotube atoms over $10$\% $L$.}
\end{figure}

\begin{figure}[ht]
\centerline{\includegraphics[width=10cm]{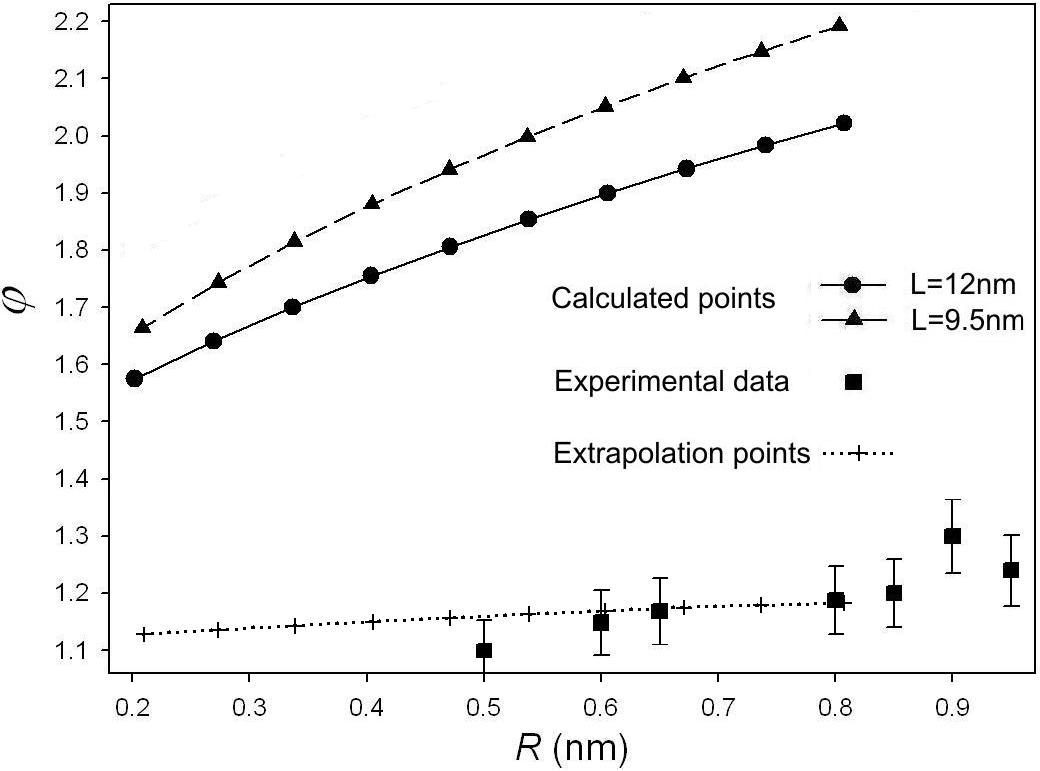}}
\caption{\label{fig:Rdependence}
Ratio of charge enhancement $\varphi$ for a number of tubes with different radius $R$, on a SiO$_{2}$ surface. The solid squares stand for $\varphi$ derived from the experimental measurements of seven nanotubes
(with lenght $L =1 \thicksim 9$ $\mu$m) deposited on $200$ nm silicon oxide layer. The symbols ``$+$'' stand for the extrapolation results for $\mu$m length tubes.}
\end{figure}

Since the nanotubes used in experiments have lengths in the micrometer range, and since this scale can hardly be directly addressed by calculations using atomic models due to the limit of computational resources, the issue about the relationship between the tube length $L$ and the charge distribution needs to be carefully addressed, so as to later extrapolate charge enhancement factors to the length scales of interest in experiments. The length dependence of the charge enhancements at the nanotube ends is illustrated in Fig.\,\ref{fig:sigma6length}, in which we plotted the local average charge density as a function of the position along the nanotube (the $x$-axis origin in Fig.\,\ref{fig:sigma6length} corresponds to the nanotube midpoint). The local average charge density is defined from the charge carried by individual CNT atoms, when averaged along the nanotube circumference and along a fraction of the length $L$ of the CNT (this fraction is taken as 10\%$L$ in Fig.\,\ref{fig:sigma6length}). $\sigma^{ave}$ is the quantity which can be accessed experimentally from EFM techniques.\cite{Zdrojek2008} The typical shape of the CNT charge distribution observed in Fig.\,\ref{fig:sigma6length} corresponds to the U-like shape expected for short nanotubes, \cite{Keblinski2002} but the charge enhancement at the tube ends is already seen to become less significant when the tube gets longer. Furthermore, we can also see that the charge enhancement is weaker when the ends of the tubes are closed and when the nanotubes are placed on the SiO$_2$ surface. The latter effect can be understood by the fact that the net nanotube charge is located at the CNT side close to the substrate (see Fig.\,\ref{fig:map90tip} (c) and (d)), which leads to an effective reduction of the charge-distributed area in the non-axial direction, similar
to an effective decrease of the nanotube radius $R$, which will be discussed further in this paper (see Fig.\,\ref{fig:Rdependence}).

\begin{figure}[ht]
\centerline{\includegraphics[width=10cm]{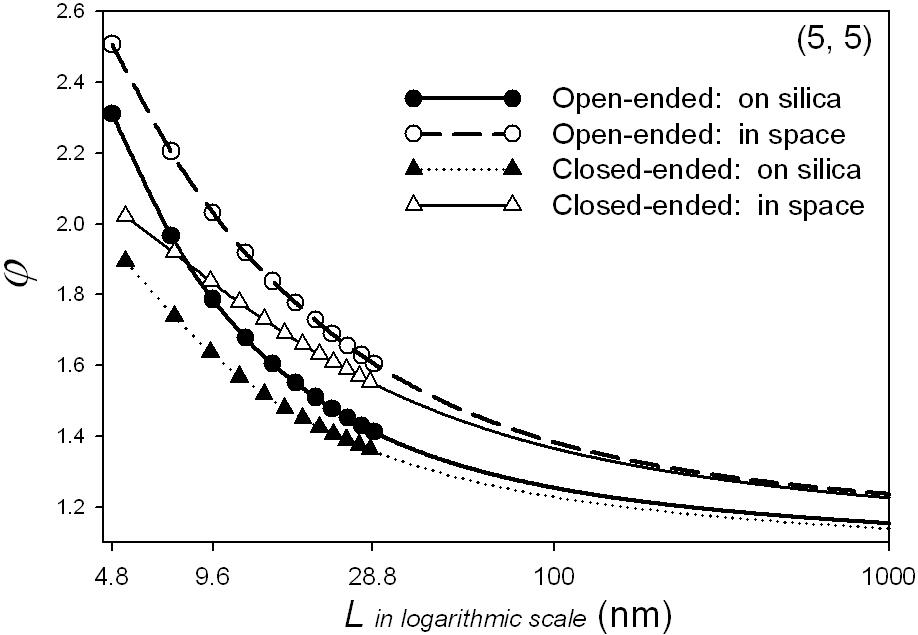}}
\caption{\label{fig:lengthfit12}
Ratio of charge enhancement $\varphi$ as a function of tube length $L$ (in common logarithmic scale).
This ratio is calculated for both an open-ended (circles) and a closed-ended (triangles) (5,5) SWCNTs (radius $R \approx 0.34$ nm) placed upon a SiO$_{2}$ surface (solid symbols) with $d = 0.34$ nm, and is compared with that for the same tubes in space (empty symbols). The symbols present the calculated points, and the lines stand for the extrapolation curves.}
\end{figure}

We now focus on charge enhancements for $\mu$m-long nanotubes, and their comparison with experimental results.
Since the spatial resolution in EFM experiments is about $100$nm (this resolution is mostly limited
by the tip-substrate separation during EFM detection), we now consider the enhancement zone in our calculation as a zone of length $10\%L$ at the tube end, and
define the charge enhancement ratio $\varphi$ as the ratio between the charge density $\sigma^{end}$ averaged in the zone of length $10\%L$ at the end of the nanotube, and
the charge density $\sigma^{middle}$ at the center of the nanotube. The influence of the tube length on the charge enhancement ratio $\varphi=\sigma^{end}/\sigma^{middle}$
is shown in Fig.\,\ref{fig:lengthfit12}. $\varphi$ is seen to decrease significantly with $L$ for short tubes (particularly for $L<10$ nm), but the variations
get smaller when the tube is longer. Note that $\varphi$ is independent of $\sigma^{ave}$, because the local charge densities should be proportional
to the total one by requiring a constant electric potential on the tube surface.
Furthermore, we find that if the (open or closed) cap structure plays an important role in the charge enhancement for short tubes ($L<15$ nm) (as seen in Fig.\,\ref{fig:map90tip} ),
this effect already becomes unsignificant for $L\approx 30$ nm, and will become negligible for $\mu$m-long nanotubes in experiments with $\approx 100nm$ resolution.
Finally, it appears that the only parameter that needs to be properly taken into account is the presence of the SiO$_2$ surface below the nanotube, which still effectively
reduces the charge enhancement ration $\varphi$ for $L=30$ nm.

In order to extrapolate these results towards $\mu$m-long nanotubes, we performed a fit of the data points of Fig.\,\ref{fig:lengthfit12}. Since the
the analytical formula of the exact distribution of charge on a hollow tube is not known in the literature, we used
the equation : $\varphi = \ln( a_{1} \times L + a_{2}) / \ln(a_{3} \times L + a_{4})$, in which $a_{n}\ (1\leq n\leq 4)$ are four fitted parameters for each
nanotube radius. This phenomenological equation has been chosen since it describes a ratio between two cylindrical capacitances,
and is thus well-suited to account for $\varphi$, which is the ratio between the linear charge densities at the end and at the middle of the nanotube.
The lines in Fig.\,\ref{fig:lengthfit12} correspond to the fits obtained independently for the
nanotubes with either open or closed caps, vacuum environment, or SiO$_2$ surface. The extrapolated values for $\varphi$ are seen to converge for large $L$ for
open and closed cap structures, but to differ depending on the vacuum or SiO$_2$ environment. This behaviour is in full agreement with the trend
observed on the atomic calculation points obtained for $L\approx30$ nm, which already brings confidence at this stage about the validity of our extrapolation procedure.

\begin{figure}[ht]
\centerline{\includegraphics[width=10cm]{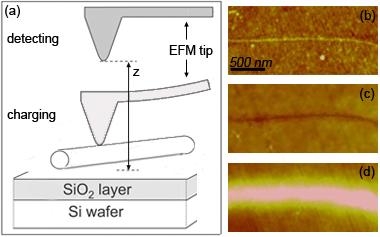}}
\caption{\label{fig:experiment1}
(Color online) (a) Schematics of the charge injection and detection with the EFM tip. Charging takes place when the biased tip is put in contact with the CNT. During the data acquisition cantilever is lifted at distance $z$ above the surface. (b) Topography image of a SWCNT with $0.5$ nm radius deposited on $200$ nm oxide layer. (c) EFM image acquired before injection. The dark feature corresponds to the uncharged tube. (d) EFM scan after injection experiment. The bright feature corresponds to the charged tube with the uniform linear charge density of $0.055e/$nm.}
\end{figure}

Our theoretical predictions are finally compared with electrostatic measurements performed by injecting and detecting charges in individual CNTs using electrostatic
force microscopy. In these experiments, nanotubes grown by chemical vapour deposition are deposited from dichloromethane solutions onto silicon wafers covered by
a 200nm-thick thermal dioxide layers. Individual nanotubes are located by atomic force microscopy, and then charged (see Fig.\,\ref{fig:experiment1} (a))
by pressing the biased tip of an atomic force microscope on the nanotube (typically with an injection bias $V_{inj}=-5$V, pressing force of a few nN).
The CNT charge state is then measured before and after injection by EFM, by
recording electrostatic force gradients acting on the tip which is intentionally lifted at a distance $z$ about $50$-$100$ nm
above the sample surface to discard short range surface forces. Fig.\,\ref{fig:experiment1} (b) shows the topography image of a SWCNT.
In Fig.\,\ref{fig:experiment1} (c), the EFM scan of the tube before charging is shown, as a dark footprint of the CNT topography
associated with attractive forces due to the nanotube capacitance. It can be shown experimentally that the negative frequency shifts
are here of capacitive origin, and not originating in a positive charge transferred from the substrate to the nanotube
(see details in Ref. \cite{Zdrojek2006}). The nanotube EFM image after charge injection is shown in Fig.\,\ref{fig:experiment1} (d).
The tube is seen here as a bright feature as a result of the negative charges injected in the tube.
From previous EFM studies, we have shown that the charge imaged for SWCNTs mainly correspond to charge emitted from the tube and
\lq\lq printed\rq\rq ~in the oxide layer in the vicinity of the nanotube.\cite{Zdrojek20062,Paillet2006}

\begin{figure}[ht]
\centerline{\includegraphics[width=8cm]{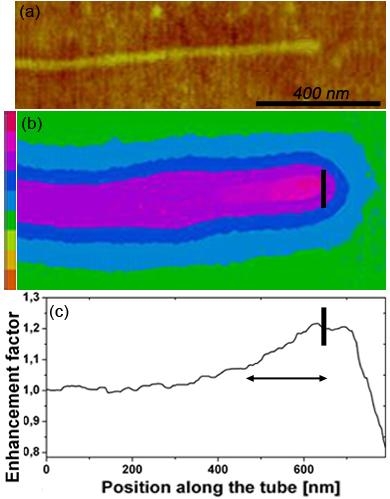}}
\caption{\label{fig:experiment2}
(Color online) (a) AFM topography image of single nanotube with $R=0.8$ nm and $L\approx 2$ $\mu$m deposited on silicon dioxide.
(b) EFM scan of the same tube made after charge injection. A non-linear color scale has been used in order to clearly show  the weak enhancement at the tube end.
The black line is a guide-to-the-eye for the physical end of the tube. (c) Experimental charge enhancement along the axis of the nanotube,
defined as the ratio of the EFM signal with that measured at the middle of the nanotube.}
\end{figure}


To compare these predictions with our calculation results, we show in Fig.\,\ref{fig:experiment2} the charge distribution at the end of a SWCNT (total length 2 $\mu$m)
after a charge injection experiment. A non-linear color scale has been used in Fig.\,\ref{fig:experiment2}b in order to evidence the weak charge enhancement
localized within $200$ nm at the nanotube end. The charge distribution along the nanotube is shown in Fig.\,\ref{fig:experiment2}c, in which we plotted the
charge enhancement factor measured from EFM, defined as the ratio of the EFM signal with that measured at the middle of the nanotube. From these experimental data, one gets the maximum value $\varphi = 1.17\pm 0.05$ for this tube (see Fig.\,\ref{fig:experiment2} (c)),
in agreement with the numerical extrapolation from theoretical results predicting $\varphi \approx 1.165$ for a $2$ $\mu$m tube with $R=0.8$ nm
deposited on a SiO$_2$ surface.

To further validate the comparison of our theoretical predictions with experiments, we now focus on the dependence of the charge enhancement ratio $\varphi$ as a function of the nanotube radius $R$. Such an analysis would not be possible for short nanotubes, because the charge enhancement ratio would then be strongly dependent on the nanotube cap structure, as discussed previously (see Fig.\,\ref{fig:lengthfit12}), while this effect is not relevant for $\mu$m-long naontubes. Intuitively, one can guess that the nanotube charge enhancement factor will increase with the tube radius $R$, because the enhancement factor decreases with the nanotube length $L$ : increasing $R$ at fixed length $L$ reduces the nanotube anisotropy, and is qualitatively similar as decreasing the nanotube length $L$ for a fixed radius $R$. 

Experimentally, we measured the charge densities along seven SWCNTs with lengths between 1 and 9 $\mu$m in a similar way as in Fig.\,\ref{fig:experiment2} (c), and plotted the corresponding charge enhancement ratios in Fig.\,\ref{fig:Rdependence}, as a function of the nanotube radius $R$ measured from atomic force topography images. The $\pm 0.05$ error bars on $\varphi$ correspond here to the accuracy of the EFM measurements. Experimental data points clearly show that $\varphi$ slightly increases as a function of the nanotube radius $R$. The possibility to observe this behavior also confirms that the values of $\varphi$ on $\mu$m-long nanotubes do not critically depend on the tube length, nor on the nanotube cap structure. Numerical calculations for the charge enhancement ratio $\varphi$ have also been performed using nanotubes with different radius $R$, and are shown in Fig.\,\ref{fig:Rdependence}. Direct calculations of $\varphi$ obtained from the atomic dipole-charge models and using an averaging over 10\% $L$ are given in Fig.\,\ref{fig:Rdependence} for two short nanotubes (9 and 12 nm, solid circles and triangles), as well as calculation results obtained for $\mu$m-long nanotubes (dotted line) using the extrapolation procedure described in Fig.\,\ref{fig:lengthfit12}. Theoretical predictions are seen to quantitatively agree with experimental data within experimental error bars, and confirm the increase of the charge enhancement ratio $\varphi$ as a function of the nanotube radius. The values of $\varphi$ computed with 5\%$L$ and 15\%$L$ for $\mu$m length tubes (data not shown) also vary within experimental error bars.

In summary, we have characterized the enhancement of net electric charge in SWCNTs by both atomic-model calculations and EFM experiments.
We have demonstrated that the U-like shape of the charge distribution expected for short nanotubes is replaced for $\mu$m-long nanotubes by weak charge enhancements
localized at the nanotube ends, while the nanotube charge densities are otherwise almost constant along the nanotubes.
The dependence on the tube length, nanotube cap structure, and the influence of silica substrate have been investigated.
It has been shown that the charge enhancement at the ends of CNTs depends strongly on the geometry of the cap only for short tubes ($<100$ nm), but has an
insignificant influence for nanotubes with lengths in the micrometer range. The increase of the charge enhancement ratio with the nanotube radius has been
demonstrated experimentally, in quantitative agreement with theoretical predictions. 

We expect that the mapping and the understanding of the charge enhancement of CNTs are important for many applications,
besides the fundamental character of this study, e.g. imaging of field and charge in CNTs electronic circuits\cite{Bachtold2000,De2001}
or emission devices modified by the presence of surfaces.
The electrostatic response of nanotubes appears to be strongly sensitive to its environment,
which is of high importance for nanotube based sensors.\cite{Snow2005,Robinson2006}
This work can also have implication in the field of nano-electromechanical systems and charge storage devices.

This work is a part of the CNRS GDR-E No. 2756. Z. W. acknowledges the support from the region of Franche-Comt\'{e} (grant 060914-10).


\begin{thebibliography}{24}
\expandafter\ifx\csname natexlab\endcsname\relax\def\natexlab#1{#1}\fi
\expandafter\ifx\csname bibnamefont\endcsname\relax
  \def\bibnamefont#1{#1}\fi
\expandafter\ifx\csname bibfnamefont\endcsname\relax
  \def\bibfnamefont#1{#1}\fi
\expandafter\ifx\csname citenamefont\endcsname\relax
  \def\citenamefont#1{#1}\fi
\expandafter\ifx\csname url\endcsname\relax
  \def\url#1{\texttt{#1}}\fi
\expandafter\ifx\csname urlprefix\endcsname\relax\def\urlprefix{URL }\fi
\providecommand{\bibinfo}[2]{#2}
\providecommand{\eprint}[2][]{\url{#2}}

\bibitem[{\citenamefont{Anantram and Leonard}(2006)}]{Anantram2006}
\bibinfo{author}{\bibfnamefont{M.}~\bibnamefont{Anantram}} \bibnamefont{and}
  \bibinfo{author}{\bibfnamefont{F.}~\bibnamefont{Leonard}},
  \bibinfo{journal}{Rep. Prog. Phys.} \textbf{\bibinfo{volume}{69}},
  \bibinfo{pages}{507} (\bibinfo{year}{2006}).

\bibitem[{\citenamefont{De~Heer et~al.}(1995)\citenamefont{De~Heer, Chatelain,
  and Ugarte}}]{De1995}
\bibinfo{author}{\bibfnamefont{W.}~\bibnamefont{De~Heer}},
  \bibinfo{author}{\bibfnamefont{A.}~\bibnamefont{Chatelain}},
  \bibnamefont{and} \bibinfo{author}{\bibfnamefont{D.}~\bibnamefont{Ugarte}},
  \bibinfo{journal}{Science} \textbf{\bibinfo{volume}{270}},
  \bibinfo{pages}{1179} (\bibinfo{year}{1995}).

\bibitem[{\citenamefont{Snow et~al.}(2005)\citenamefont{Snow, Perkins, Houser,
  Badescu, and Reinecke}}]{Snow2005}
\bibinfo{author}{\bibfnamefont{E.}~\bibnamefont{Snow}},
  \bibinfo{author}{\bibfnamefont{F.}~\bibnamefont{Perkins}},
  \bibinfo{author}{\bibfnamefont{E.}~\bibnamefont{Houser}},
  \bibinfo{author}{\bibfnamefont{S.}~\bibnamefont{Badescu}}, \bibnamefont{and}
  \bibinfo{author}{\bibfnamefont{T.}~\bibnamefont{Reinecke}},
  \bibinfo{journal}{Science} \textbf{\bibinfo{volume}{307}},
  \bibinfo{pages}{1942} (\bibinfo{year}{2005}).

\bibitem[{\citenamefont{An et~al.}(2001)\citenamefont{An, Kim, Park, Choi, Lee,
  Chung, Bae, Lim, and Lee}}]{An2001}
\bibinfo{author}{\bibfnamefont{K.}~\bibnamefont{An}},
  \bibinfo{author}{\bibfnamefont{W.}~\bibnamefont{Kim}},
  \bibinfo{author}{\bibfnamefont{Y.}~\bibnamefont{Park}},
  \bibinfo{author}{\bibfnamefont{Y.}~\bibnamefont{Choi}},
  \bibinfo{author}{\bibfnamefont{S.}~\bibnamefont{Lee}},
  \bibinfo{author}{\bibfnamefont{D.}~\bibnamefont{Chung}},
  \bibinfo{author}{\bibfnamefont{D.}~\bibnamefont{Bae}},
  \bibinfo{author}{\bibfnamefont{S.}~\bibnamefont{Lim}}, \bibnamefont{and}
  \bibinfo{author}{\bibfnamefont{Y.}~\bibnamefont{Lee}}, \bibinfo{journal}{Adv.
  Mater.} \textbf{\bibinfo{volume}{13}}, \bibinfo{pages}{497}
  (\bibinfo{year}{2001}).

\bibitem[{\citenamefont{Cui et~al.}(2002)\citenamefont{Cui, Sordan, Burghard,
  and Kern}}]{Cui2002}
\bibinfo{author}{\bibfnamefont{J.}~\bibnamefont{Cui}},
  \bibinfo{author}{\bibfnamefont{R.}~\bibnamefont{Sordan}},
  \bibinfo{author}{\bibfnamefont{M.}~\bibnamefont{Burghard}}, \bibnamefont{and}
  \bibinfo{author}{\bibfnamefont{K.}~\bibnamefont{Kern}},
  \bibinfo{journal}{Appl. Phys. Lett.} \textbf{\bibinfo{volume}{81}},
  \bibinfo{pages}{3260} (\bibinfo{year}{2002}).

\bibitem[{\citenamefont{Ryu et~al.}(2007)\citenamefont{Ryu, Huang, and
  Choi}}]{Ryu2007}
\bibinfo{author}{\bibfnamefont{S.-W.} \bibnamefont{Ryu}},
  \bibinfo{author}{\bibfnamefont{X.-J.} \bibnamefont{Huang}}, \bibnamefont{and}
  \bibinfo{author}{\bibfnamefont{Y.-K.} \bibnamefont{Choi}},
  \bibinfo{journal}{Appl. Phys. Lett.} \textbf{\bibinfo{volume}{91}},
  \bibinfo{pages}{063110} (\bibinfo{year}{2007}).

\bibitem[{\citenamefont{Bockrath et~al.}(2002)\citenamefont{Bockrath, Markovic,
  Shepard, Tinkham, Gurevich, Kouwenhoven, Wu, and Sohn}}]{Bockrath2002}
\bibinfo{author}{\bibfnamefont{M.}~\bibnamefont{Bockrath}},
  \bibinfo{author}{\bibfnamefont{N.}~\bibnamefont{Markovic}},
  \bibinfo{author}{\bibfnamefont{A.}~\bibnamefont{Shepard}},
  \bibinfo{author}{\bibfnamefont{M.}~\bibnamefont{Tinkham}},
  \bibinfo{author}{\bibfnamefont{L.}~\bibnamefont{Gurevich}},
  \bibinfo{author}{\bibfnamefont{L.}~\bibnamefont{Kouwenhoven}},
  \bibinfo{author}{\bibfnamefont{M.}~\bibnamefont{Wu}}, \bibnamefont{and}
  \bibinfo{author}{\bibfnamefont{L.}~\bibnamefont{Sohn}},
  \bibinfo{journal}{Nano Letters} \textbf{\bibinfo{volume}{2}},
  \bibinfo{pages}{187} (\bibinfo{year}{2002}).

\bibitem[{\citenamefont{Jespersen and Nygard}(2007)}]{Jespersen2007}
\bibinfo{author}{\bibfnamefont{T.}~\bibnamefont{Jespersen}} \bibnamefont{and}
  \bibinfo{author}{\bibfnamefont{J.}~\bibnamefont{Nygard}},
  \bibinfo{journal}{Appl. Phys. A} \textbf{\bibinfo{volume}{88}},
  \bibinfo{pages}{309} (\bibinfo{year}{2007}).

\bibitem[{\citenamefont{Paillet et~al.}(2005)\citenamefont{Paillet, Poncharal,
  and Zahab}}]{Paillet2005}
\bibinfo{author}{\bibfnamefont{M.}~\bibnamefont{Paillet}},
  \bibinfo{author}{\bibfnamefont{P.}~\bibnamefont{Poncharal}},
  \bibnamefont{and} \bibinfo{author}{\bibfnamefont{A.}~\bibnamefont{Zahab}},
  \bibinfo{journal}{Phys.\ Rev. Lett.} \textbf{\bibinfo{volume}{94}},
  \bibinfo{pages}{186801} (\bibinfo{year}{2005}).

\bibitem[{\citenamefont{Zdrojek
  et~al.}(2006{\natexlab{a}})\citenamefont{Zdrojek, M\'{e}lin, Diesinger,
  Sti\'{e}venard, Gebicki, and Adamowicz}}]{Zdrojek2006}
\bibinfo{author}{\bibfnamefont{M.}~\bibnamefont{Zdrojek}},
  \bibinfo{author}{\bibfnamefont{T.}~\bibnamefont{M\'{e}lin}},
  \bibinfo{author}{\bibfnamefont{H.}~\bibnamefont{Diesinger}},
  \bibinfo{author}{\bibfnamefont{D.}~\bibnamefont{Sti\'{e}venard}},
  \bibinfo{author}{\bibfnamefont{W.}~\bibnamefont{Gebicki}}, \bibnamefont{and}
  \bibinfo{author}{\bibfnamefont{L.}~\bibnamefont{Adamowicz}},
  \bibinfo{journal}{J. Appl. Phys.} \textbf{\bibinfo{volume}{100}},
  \bibinfo{pages}{114326} (\bibinfo{year}{2006}{\natexlab{a}}).

\bibitem[{\citenamefont{Keblinski et~al.}(2002)\citenamefont{Keblinski, Nayak,
  Zapol, and Ajayan}}]{Keblinski2002}
\bibinfo{author}{\bibfnamefont{P.}~\bibnamefont{Keblinski}},
  \bibinfo{author}{\bibfnamefont{S.}~\bibnamefont{Nayak}},
  \bibinfo{author}{\bibfnamefont{P.}~\bibnamefont{Zapol}}, \bibnamefont{and}
  \bibinfo{author}{\bibfnamefont{P.}~\bibnamefont{Ajayan}},
  \bibinfo{journal}{Phys.\ Rev. Lett.} \textbf{\bibinfo{volume}{89}},
  \bibinfo{pages}{255503} (\bibinfo{year}{2002}).

\bibitem[{\citenamefont{Li and Chou}(2006)}]{Li2006a}
\bibinfo{author}{\bibfnamefont{C.}~\bibnamefont{Li}} \bibnamefont{and}
  \bibinfo{author}{\bibfnamefont{T.-W.} \bibnamefont{Chou}},
  \bibinfo{journal}{Appl. Phys. Lett.} \textbf{\bibinfo{volume}{89}},
  \bibinfo{pages}{063103} (\bibinfo{year}{2006}).

\bibitem[{\citenamefont{Mayer}(2007)}]{mayer-07-01}
\bibinfo{author}{\bibfnamefont{A.}~\bibnamefont{Mayer}},
  \bibinfo{journal}{Phys.\ Rev.~B} \textbf{\bibinfo{volume}{75}},
  \bibinfo{pages}{045407} (\bibinfo{year}{2007}).

\bibitem[{\citenamefont{Mayer and Astrand}(2008)}]{Mayer2008}
\bibinfo{author}{\bibfnamefont{A.}~\bibnamefont{Mayer}} \bibnamefont{and}
  \bibinfo{author}{\bibfnamefont{P.-O.} \bibnamefont{Astrand}},
  \bibinfo{journal}{J. Phys. Chem. A} \textbf{\bibinfo{volume}{112}},
  \bibinfo{pages}{1277} (\bibinfo{year}{2008}).

\bibitem[{\citenamefont{Jackson}(1975)}]{Jacksonbook1975}
\bibinfo{author}{\bibfnamefont{J.~D.} \bibnamefont{Jackson}},
  \emph{\bibinfo{title}{Classical Electrodynamics}} (\bibinfo{publisher}{Wiley,
  New York}, \bibinfo{year}{1975}), \bibinfo{note}{p. 54-62}.

\bibitem[{\citenamefont{Tsetseris and Pantelides}(2006)}]{Tsetseris2006}
\bibinfo{author}{\bibfnamefont{L.}~\bibnamefont{Tsetseris}} \bibnamefont{and}
  \bibinfo{author}{\bibfnamefont{S.}~\bibnamefont{Pantelides}},
  \bibinfo{journal}{Phys.\ Rev. Lett.} \textbf{\bibinfo{volume}{97}},
  \bibinfo{pages}{266805} (\bibinfo{year}{2006}).

\bibitem[{\citenamefont{Wojdel and Bromley}(2005)}]{Wojdel2005}
\bibinfo{author}{\bibfnamefont{J.}~\bibnamefont{Wojdel}} \bibnamefont{and}
  \bibinfo{author}{\bibfnamefont{S.}~\bibnamefont{Bromley}},
  \bibinfo{journal}{J. Phys. Chem. B} \textbf{\bibinfo{volume}{109}},
  \bibinfo{pages}{1387} (\bibinfo{year}{2005}).

\bibitem[{\citenamefont{Stuart et~al.}(2000)\citenamefont{Stuart, Tutein, and
  Harrison}}]{stuart-00}
\bibinfo{author}{\bibfnamefont{S.~J.} \bibnamefont{Stuart}},
  \bibinfo{author}{\bibfnamefont{A.~B.} \bibnamefont{Tutein}},
  \bibnamefont{and} \bibinfo{author}{\bibfnamefont{J.~A.}
  \bibnamefont{Harrison}}, \bibinfo{journal}{J. Chem. Phys.}
  \textbf{\bibinfo{volume}{112}}, \bibinfo{pages}{6472} (\bibinfo{year}{2000}).

\bibitem[{\citenamefont{Zdrojek et~al.}(2008)\citenamefont{Zdrojek, Heim,
  Brunel, Mayer, and M\'{e}lin}}]{Zdrojek2008}
\bibinfo{author}{\bibfnamefont{M.}~\bibnamefont{Zdrojek}},
  \bibinfo{author}{\bibfnamefont{T.}~\bibnamefont{Heim}},
  \bibinfo{author}{\bibfnamefont{D.}~\bibnamefont{Brunel}},
  \bibinfo{author}{\bibfnamefont{A.}~\bibnamefont{Mayer}}, \bibnamefont{and}
  \bibinfo{author}{\bibfnamefont{T.}~\bibnamefont{M\'{e}lin}},
  \bibinfo{journal}{Phys.\ Rev.~B} \textbf{\bibinfo{volume}{77}},
  \bibinfo{pages}{033404} (\bibinfo{year}{2008}).


\bibitem[{\citenamefont{Zdrojek
  et~al.}(2006{\natexlab{b}})\citenamefont{Zdrojek, M\'{e}lin, Diesinger,
  Sti\'{e}venard, Gebicki, Adamowicz}}]{Zdrojek20062}
\bibinfo{author}{\bibfnamefont{M.}~\bibnamefont{Zdrojek}},
  \bibinfo{author}{\bibfnamefont{T.}~\bibnamefont{M\'{e}lin}},
  \bibinfo{author}{\bibfnamefont{H.}~\bibnamefont{Diesinger}},
  \bibinfo{author}{\bibfnamefont{D.}~\bibnamefont{Sti\'{e}venard}},
  \bibinfo{author}{\bibfnamefont{W.}~\bibnamefont{Gebicki}},
  \bibinfo{author}{\bibfnamefont{L.}~\bibnamefont{Adamowicz}},
  \bibinfo{journal}{Phys.\ Rev. Lett.} \textbf{\bibinfo{volume}{96}},
  \bibinfo{pages}{039703} (\bibinfo{year}{2006}{\natexlab{b}}).
  
\bibitem[{\citenamefont{Paillet et~al.}(2005)\citenamefont{Paillet, Poncharal,
  and Zahab}}]{Paillet2006}
\bibinfo{author}{\bibfnamefont{M.}~\bibnamefont{Paillet}},
  \bibinfo{author}{\bibfnamefont{P.}~\bibnamefont{Poncharal}},
  \bibnamefont{and} \bibinfo{author}{\bibfnamefont{A.}~\bibnamefont{Zahab}},
  \bibinfo{journal}{Phys.\ Rev. Lett.} \textbf{\bibinfo{volume}{96}},
  \bibinfo{pages}{039704} (\bibinfo{year}{2006}).
  
\bibitem[{\citenamefont{Bachtold et~al.}(2000)\citenamefont{Bachtold, Fuhrer,
  Plyasunov, Forero, Andersen, Zettl, and McEuen}}]{Bachtold2000}
\bibinfo{author}{\bibfnamefont{A.}~\bibnamefont{Bachtold}},
  \bibinfo{author}{\bibfnamefont{M.}~\bibnamefont{Fuhrer}},
  \bibinfo{author}{\bibfnamefont{S.}~\bibnamefont{Plyasunov}},
  \bibinfo{author}{\bibfnamefont{M.}~\bibnamefont{Forero}},
  \bibinfo{author}{\bibfnamefont{E.}~\bibnamefont{Andersen}},
  \bibinfo{author}{\bibfnamefont{A.}~\bibnamefont{Zettl}}, \bibnamefont{and}
  \bibinfo{author}{\bibfnamefont{P.}~\bibnamefont{McEuen}},
  \bibinfo{journal}{Phys.\ Rev. Lett.} \textbf{\bibinfo{volume}{84}},
  \bibinfo{pages}{6082} (\bibinfo{year}{2000}).

\bibitem[{\citenamefont{De~Pablo et~al.}(2001)\citenamefont{De~Pablo,
  Gomez-Navarro, Gil, Colchero, Martinez, Benito, Maser, Gomez-Herrero, and
  Baro}}]{De2001}
\bibinfo{author}{\bibfnamefont{P.}~\bibnamefont{De~Pablo}},
  \bibinfo{author}{\bibfnamefont{C.}~\bibnamefont{Gomez-Navarro}},
  \bibinfo{author}{\bibfnamefont{A.}~\bibnamefont{Gil}},
  \bibinfo{author}{\bibfnamefont{J.}~\bibnamefont{Colchero}},
  \bibinfo{author}{\bibfnamefont{M.}~\bibnamefont{Martinez}},
  \bibinfo{author}{\bibfnamefont{A.}~\bibnamefont{Benito}},
  \bibinfo{author}{\bibfnamefont{W.}~\bibnamefont{Maser}},
  \bibinfo{author}{\bibfnamefont{J.}~\bibnamefont{Gomez-Herrero}},
  \bibnamefont{and} \bibinfo{author}{\bibfnamefont{A.}~\bibnamefont{Baro}},
  \bibinfo{journal}{Appl. Phys. Lett.} \textbf{\bibinfo{volume}{79}},
  \bibinfo{pages}{2979} (\bibinfo{year}{2001}).

\bibitem[{\citenamefont{Robinson et~al.}(2006)\citenamefont{Robinson, Snow,
  Badescu, Reinecke, and Perkins}}]{Robinson2006}
\bibinfo{author}{\bibfnamefont{J.}~\bibnamefont{Robinson}},
  \bibinfo{author}{\bibfnamefont{E.}~\bibnamefont{Snow}},
  \bibinfo{author}{\bibfnamefont{S.}~\bibnamefont{Badescu}},
  \bibinfo{author}{\bibfnamefont{T.}~\bibnamefont{Reinecke}}, \bibnamefont{and}
  \bibinfo{author}{\bibfnamefont{F.}~\bibnamefont{Perkins}},
  \bibinfo{journal}{Nano Letters} \textbf{\bibinfo{volume}{6}},
  \bibinfo{pages}{1747} (\bibinfo{year}{2006}).

\end{thebibliography}
\end{document}